\documentclass[3p,times]{elsarticle}
\usepackage[framed,numbered,autolinebreaks,useliterate]{mcode}
\usepackage{booktabs}
\usepackage{multirow}

\usepackage{enumitem}
\newlist{todolist}{itemize}{2}
\setlist[todolist]{label=$\square$}
\usepackage{pifont}
\usepackage{amssymb}
\usepackage{amsmath}

\usepackage[figuresright]{rotating}

\usepackage{subfig}

\usepackage{graphicx}
\usepackage{dcolumn}
\usepackage{bm}
\usepackage{epstopdf}
\usepackage{epsfig}
\usepackage{xcolor}
\colorlet{red}{black}
\colorlet{blue}{black}
\definecolor{proofred}{RGB}{255,0,0}

\usepackage[colorlinks = true,
linkcolor = blue,
urlcolor  = blue,
citecolor = blue,
anchorcolor = blue]{hyperref}
\usepackage{url}

\DeclareMathOperator{\sech}{sech}

\begin{document}

\begin{frontmatter}
	
	
	
	
	\title{{Multiprecision computation of bright and dark solitons in the discrete nonlinear Schr\"odinger equation}
	}
	
	\author[au2,au3]{Rudy Kusdiantara \corref{cor1}}
	\ead{rudy\_kusdiantara@itb.ac.id}
	\author[au1]{Farrell T.\ Adriano}
	\author[au1]{Hadi Susanto}
	\cortext[cor1]{Corresponding author}
	\address[au2]{Industrial and Financial Mathematics Research Group, Institut Teknologi Bandung, Jl.\ Ganesha No.\ 10, Bandung, 40132, Indonesia}
	\address[au3]{Center of Excellence in Predictive Risk and Simulation Modeling, Institut Teknologi Bandung, Jl.\ Ganesha No.\ 10, Bandung, 40132, Indonesia}
	\address[au1]{Department of Mathematics, Khalifa University, PO Box 127788, Abu Dhabi, United Arab Emirates}

	\begin{abstract}		
		{
We study the spectral stability of bright and dark solitons in the discrete nonlinear Schr\"odinger (DNLS) equation using multiprecision arithmetic. The eigenvalues governing stability are exponentially small in the lattice spacing and cannot be resolved with standard double precision. To address this, we develop a computational framework combining multiprecision arithmetic, an exact Jacobian for the stationary problem, and a squared-operator formulation for spectral analysis. This enables accurate resolution of exponentially small eigenvalues and direct comparison with exponential-asymptotic predictions. Our results show that onsite bright solitons are spectrally stable, whereas intersite bright solitons and both onsite and intersite dark solitons are unstable. Bright solitons require only a few eigenvalues and allow efficient large-scale computations, while dark solitons demand higher precision due to their proximity to the continuous spectrum. Simulations up to \(N=65{,}250\) grid points (31.7 GB RAM) highlight the necessity of multiprecision arithmetic for capturing beyond-all-orders spectral effects.
		}
	\end{abstract}

	
\end{frontmatter}


\section{Introduction}
Nonlinear waves in discrete media arise naturally in many branches of physics, including optics \cite{maimistov1999nonlinear}, condensed matter systems \cite{fitzmaurice1993nonlinear}, and Bose--Einstein condensates \cite{kevrekidis2008emergent}. A central model is the discrete nonlinear Schr\"odinger (DNLS) equation, which generalizes the continuous nonlinear Schr\"odinger (NLS) equation to a lattice setting \cite{Kevrekidis2009,KivsharAgrawal2003}. The DNLS supports both bright and dark solitons depending on the sign of the nonlinearity, with additional richness compared to the continuous case due to the breaking of translation invariance. In particular, two fundamental families of solutions emerge: \emph{onsite solitons}, centered on lattice sites, and \emph{intersite solitons}, centered between lattice sites. Their stability properties are subtle and strongly affected by exponentially small eigenvalues that standard asymptotics or double-precision computations cannot resolve. \textcolor{red}{Here and throughout, a quantity is said to be \emph{exponentially small} in the lattice spacing $h$ if it scales as $\mathcal{O}\!\left(e^{-c/h^p}\right)$ for some constants $c>0$ and $p>0$ as $h\to0^+$. Throughout the remainder of the paper, $h\to0$ is understood to mean $h\to0^+$.
Such terms decay faster than any power of $h$ and therefore lie beyond all orders of a standard algebraic asymptotic expansion in $h$; capturing them requires exponential (beyond-all-orders) asymptotics.}

The study of stability in DNLS solitons has a long history. Early works by Kapitula and co-authors rigorously analyzed the existence and stability of bright and dark lattice solitons, and emphasized the role of small eigenvalues in determining their dynamics \cite{KapitulaKevrekidis2001Nonlinearity,Kapitula2001PhysicaD,KapitulaKevrekidisSandstede2004}. These studies established that the fate of onsite and intersite configurations depends delicately on exponentially small interactions between lattice modes. More recent progress has been made using \emph{exponential asymptotics}, which systematically captures beyond-all-orders effects responsible for soliton selection and stability. Adriano \emph{et al}.\ \cite{Adriano2025} developed this approach for the DNLS, showing analytically that onsite bright solitons are spectrally stable, while intersite bright solitons and both branches of dark solitons are unstable. Their asymptotic predictions were in good agreement with high-precision numerics, but they also highlighted a major computational challenge: the relevant eigenvalues are exponentially small in the lattice spacing and therefore invisible to standard double-precision arithmetic. In complementary work, Lustri, Kevrekidis, and Chapman \cite{Lustri2025} derived higher-order exponential-asymptotic corrections for translational modes in the focusing DNLS. Their analysis refines the prefactors and provides small-$h$ corrections to the leading exponential laws, improving quantitative agreement with numerical results and guiding the rescaled comparisons used later in this paper. \textcolor{red}{Most recently, Lustri, Kevrekidis, and Pelinovsky~\cite{LustriDark2026} obtained precise leading- and next-order exponential asymptotics for the translational eigenvalues of intersite and onsite dark solitons. Their intersite formulas provide a direct analytical benchmark for the multiprecision data reported below.}
Related ideas on exponential asymptotics have also been developed for quantum droplets and bubbles in a quadratic-cubic DNLS setting~\cite{adriano2025exponential}.

A growing body of evidence shows that resolving exponentially small effects and eigenvalues near the continuous spectrum often requires arithmetic precision beyond the standard double format.  
Classical analyses of rounding error and ill-conditioning demonstrate how tiny spectral gaps can be overwhelmed by roundoff noise~\cite{Higham2002,trefethen2020spectra}.  
\textcolor{red}{Finite-dimensional approximations of eigenvalue problems near the essential spectrum may also exhibit \emph{spectral pollution}~\cite{DaviesPlum2004}. This is an approximation effect distinct from floating-point roundoff and is not removed merely by increasing the arithmetic precision. In the present computations, multiprecision is used to suppress roundoff, whereas convergence under changes in the domain size, the number of requested eigenvalues, and the spectral window is used to assess spectral resolution.}  
Exponentially small phenomena arise in systems where dispersion or weak nonlocality produces radiative tails that are beyond all algebraic orders of standard asymptotic expansions~\cite{boyd2012weakly}; see also, for example, recent analyses of the Korteweg--de~Vries (KdV) equation with a fifth-order derivative term~\cite{fodor2023higher,fodor2024new}.  
From the computational perspective, modern arbitrary-precision libraries and toolboxes now make such studies feasible at scale~\cite{Bailey1995,BaileyBorwein2015,Johansson2017,Advanpix}, enabling systematic exploration of the trade-offs between accuracy and cost in problems where the relevant eigenvalues or spectral splittings are exponentially small.
\textcolor{red}{Nevertheless, to the best of our knowledge, no systematic study exists on how the exponentially small stability eigenvalues of lattice solitons can be computed reliably, what numerical precision and resources such computations require, and to what extent the available exponential-asymptotic predictions~\cite{Adriano2025,Lustri2025,LustriDark2026} are quantitatively confirmed or challenged by the numerics.}
This motivates the present work.


{We revisit the stability of DNLS bright and dark solitons using multiprecision arithmetic, implemented with the Advanpix Multiprecision Computing Toolbox for \textsc{Matlab} \cite{Advanpix}.
\textcolor{red}{The contribution of this work is not limited to the use of a software package; rather, we add to the existing body of literature in the following ways.
(i)~We develop a systematic and reproducible computational framework that combines multiprecision arithmetic with an exact Jacobian in the Newton iteration and a squared-operator formulation for the spectral problem, which significantly improves the robustness and accuracy of exponentially small eigenvalue calculations.
(ii)~Using this framework, we substantially extend the numerical verification of the exponential-asymptotic predictions of Adriano \emph{et al.}~\cite{Adriano2025} and of the higher-order corrections of Lustri \emph{et al.}~\cite{Lustri2025} for bright solitons, resolving eigenvalues as small as $|\lambda|\sim10^{-136}$, far beyond the range accessible in the earlier numerical comparisons and beyond the reach of double precision.
(iii)~For intersite dark solitons, we directly compare the multiprecision data with the leading-order prediction of Adriano \emph{et al.}~\cite{Adriano2025}, using the factor-of-two correction to their published expression, and with the {formula} of Lustri \emph{et al.}~\cite{LustriDark2026}. The corrected Adriano prefactor $265.913$ {is essentially identical to the leading-order term in the Lustri formula,} $266.004$, and both agree quantitatively with the numerical prefactor approaching $\approx263$, while the {full next-order formula} captures the observed finite-$h$ trend without requiring an empirical fit.
(iv)~As an independent cross-check, we analyze an associated parabolic problem, whose linearization operator possesses a spectral gap; for this problem we derive a new exponential-asymptotic approximation of the critical eigenvalue (\ref{app:parabolic}) and show that it also agrees quantitatively with the multiprecision numerics, reinforcing confidence in the combined multiprecision-plus-exponential-asymptotics approach for a second, independent class of eigenvalue problems.
(v)~We quantify the computational cost, in terms of runtime, memory, and required precision, of resolving such exponentially small spectra for the different soliton branches.}
	The \textsc{Matlab} implementation provided in \ref{app:code} illustrates this workflow in a reproducible manner. Beyond the DNLS equation, the methodology developed here is applicable to nonlinear wave problems where exponentially small spectral effects determine stability, and where standard double precision fails to provide reliable results.}

The paper is organized as follows. In Section~\ref{sec:model} we present the mathematical formulation of the DNLS equation, the construction of bright and dark soliton solutions, and their discretization. Section~\ref{sec:numerics} describes the numerical setup, including the multiprecision implementation, and presents results on the stability of onsite and intersite solitons. Section~\ref{sec:conclusion} concludes with a discussion of the advantages and disadvantages of multiprecision, its current limitations, and potential applications to other nonlinear wave problems. For reproducibility, \ref{app:code} provides \textsc{Matlab} codes illustrating the computation of stationary states and their linear stability using multiprecision arithmetic.

\section{Mathematical Formulation}\label{sec:model}

We consider the one-dimensional DNLS equation \cite{Kevrekidis2009,eilbeck2003discrete}
\begin{equation}\label{eq:DNLS}
	i\,\dot{\psi}_j + \frac{\psi_{j+1}-2\psi_j+\psi_{j-1}}{h^2}
	+ \sigma |\psi_j|^2 \psi_j = 0, \qquad j\in\mathbb{Z},
\end{equation}
where $\psi_j(t)$ is a complex-valued sequence representing the wave amplitude at the lattice site $j$, and $h>0$ denotes the lattice spacing.
The parameter $\sigma$ characterizes the type of nonlinearity: $\sigma=+1$ corresponds to the focusing case, while $\sigma=-1$ corresponds to the defocusing case.

\subsection{Standing waves}
To obtain stationary solutions, we employ the standing-wave ansatz
\begin{equation}\label{eq:ansatz_discr}
	\psi_j(t) = \phi_j\, e^{-i\omega t}, \qquad \omega\in\mathbb{R},
\end{equation}
where $\boldsymbol{\phi} = (\phi_j)$ is a real-valued profile. \textcolor{red}{Throughout the paper, the boldface symbol $\boldsymbol{\phi}$ denotes the vector of lattice amplitudes, whereas the regular symbol $\phi_j$ denotes its scalar component at site $j$.} The profile satisfies the nonlinear difference equation
\[
	-\omega\,\phi_j + \frac{\phi_{j+1}-2\phi_j+\phi_{j-1}}{h^2} + \sigma \phi_j^3 = 0.
\]
Without loss of generality, we will set $|\omega|=1$ in the following. 
We perform computations on a finite interval $j\in[-N/2+1, N/2]$. 
At the boundaries, we impose homogeneous Neumann conditions, 
\[
\phi_{-N/2} = \phi_{-N/2+1}, \qquad \phi_{N/2+1} = \phi_{N/2}.
\]
The resulting nonlinear algebraic system
\begin{equation}\label{eq:Fdiscr}
	F_j(\boldsymbol{\phi}) = -\omega \phi_j +
	\frac{\phi_{j+1}-2\phi_j+\phi_{j-1}}{h^2}
	+ \sigma \phi_j^3 = 0, 
\end{equation}
is solved numerically using the \textsc{Matlab} routine \mcode{fsolve}.  
The Jacobian matrix is provided explicitly as
\begin{equation}\label{eq:Jdiscr}
	J = \frac{1}{h^2}\,\mathrm{tridiag}(1,-2,1)\;-\;\omega I
	\;+\; 3\sigma\,\mathrm{diag}(\phi_{-N/2+1}^2,\dots,\phi_{N/2}^2),
\end{equation}
which ensures fast and reliable convergence.

\subsection{Continuous limit: the NLS equation}

In the limit $h\to 0$, the lattice spacing becomes infinitesimal, and the DNLS \eqref{eq:DNLS} formally approaches the continuous NLS equation \cite{AblowitzSegur1981,SulemSulem1999}
\begin{equation}\label{eq:NLS}
	i\,\psi_t + \psi_{xx} + \sigma |\psi|^2 \psi = 0,
\end{equation}
where $\psi(x,t)$ is the continuous interpolation of $\psi_j(t)$ and $x_j = jh$.

\textcolor{red}{Applying the same standing-wave ansatz as in \eqref{eq:ansatz_discr}, i.e.,}
\[
\psi(x,t) = \phi(x)\, e^{-i\omega t}, \qquad \omega\in\mathbb{R},
\]
\textcolor{red}{to the NLS equation \eqref{eq:NLS}} yields the stationary NLS equation
\begin{equation}\label{eq:stat}
	-\omega\,\phi + \phi_{xx} + \sigma |\phi|^2 \phi = 0.
\end{equation}
In the focusing case $(\sigma=+1)$ with $\omega>0$, Eq.~\eqref{eq:stat} admits the well-known \emph{bright soliton} solution \cite{ZakharovShabat1972}
\begin{equation}\label{eq:bright}
	\phi(x) = \sqrt{2\omega}\,\sech\!\big(\sqrt{\omega}\,x\big),
\end{equation}
which is spatially localized and decays to zero as $|x|\to\infty$.
In contrast, for the defocusing case $(\sigma=-1)$ with $\omega<0$, one obtains the \emph{dark soliton} \cite{KivsharLutherDavies1998}
\begin{equation}\label{eq:dark}
	\phi(x) = \sqrt{-\omega}\,\tanh\!\Big(\sqrt{\tfrac{-\omega}{2}}\,x\Big),
\end{equation}
which corresponds to a localized intensity dip on a nonzero background:
\[
\phi(x)\to \pm\sqrt{-\omega}, \qquad \text{as } x\to \pm\infty.
\]
\textcolor{red}{We note that the bright and dark solitons \eqref{eq:bright}--\eqref{eq:dark} are by no means the only solutions of the continuous NLS equation; owing to its integrability, the equation also admits, among others, higher-order (multi-)solitons as well as solutions on a finite background such as the Akhmediev, Kuznetsov--Ma, and Peregrine breathers, which are prototypes of rogue waves; see, e.g., the book by Akhmediev and Ankiewicz~\cite{AkhmedievAnkiewicz1997}. In this work we restrict our attention to the fundamental bright and dark solitons, since these are the solutions whose discrete onsite and intersite counterparts are the subject of our stability analysis.}

\subsection{Linear stability analysis}
To investigate the spectral stability of the discrete standing waves, we introduce a perturbation of the form
\begin{equation}\label{eq:perturb_discr}
	\psi_j(t) = e^{-i\omega t}\Big(\phi_j + [u_j+i\,v_j]e^{\lambda t}\Big),
\end{equation}
where $u_j$ and $v_j$ are real-valued perturbation components, and $\lambda\in\mathbb{C}$ is the spectral parameter.
\textcolor{red}{Substituting the perturbation ansatz \eqref{eq:perturb_discr} into \eqref{eq:DNLS} and linearizing} around the stationary state yields the discrete eigenvalue problem
\begin{equation}\label{eq:bdg_disc}
	\begin{pmatrix}
		0 & L^{(h)}_- \\
		-\,L^{(h)}_+ & 0
	\end{pmatrix}
	\begin{pmatrix} \mathbf{u} \\ \mathbf{v} \end{pmatrix}
	= \lambda \begin{pmatrix} \mathbf{u} \\ \mathbf{v} \end{pmatrix},
\end{equation}
with the discrete operators
\begin{equation}\label{eq:Lpm_disc}
	L^{(h)}_+ = -D_{xx} + \omega I - 3\sigma\,\mathrm{diag}(\boldsymbol{\phi}^2), \qquad
	L^{(h)}_- = -D_{xx} + \omega I - \sigma\,\mathrm{diag}(\boldsymbol{\phi}^2),
\end{equation}
where $D_{xx}=\tfrac{1}{h^2}\,\mathrm{tridiag}(1,-2,1)$.
The system can be reduced to a scalar problem:
\begin{equation}\label{eq:operator_L2}
	L^{(h)}_- L^{(h)}_+ \mathbf{u} = -\lambda^2 \mathbf{u}.
\end{equation}
In the limit $h\to 0$, these discrete operators converge to their continuous counterparts
\[
	L_+ = -\partial_{xx} + \omega - 3\sigma \phi^2, \qquad
	L_- = -\partial_{xx} + \omega - \sigma \phi^2,
\]
and the corresponding spectral problem
\[
	\begin{pmatrix}
		0 & L_- \\
		-L_+ & 0
	\end{pmatrix}
	\begin{pmatrix} u \\ v \end{pmatrix}
	= \lambda \begin{pmatrix} u \\ v \end{pmatrix}
\]
recovers the well-known Bogoliubov-de Gennes formulation of the NLS linear stability problem
\cite{VakhitovKolokolov1973,Pelinovsky2011}.

\section{Numerical Setup and Results}\label{sec:numerics}

All computations are performed in \textsc{Matlab} R2024b using the Advanpix Multiprecision Computing Toolbox (version~5.4.0.16035)~\cite{Advanpix}. 
The stationary discrete system~\eqref{eq:Fdiscr} is solved with the \mcode{fsolve} routine, supplying the exact Jacobian~\eqref{eq:Jdiscr}, which significantly improves convergence speed and accuracy. 
As initial guesses, we use the analytical bright- and dark-soliton profiles~\eqref{eq:bright}-\eqref{eq:dark} sampled on the computational grid. 

To analyze spectral stability, we compute the eigenvalue spectrum of the discrete operators~\eqref{eq:bdg_disc}. Instead of solving the full block eigenvalue problem, we employ the equivalent squared formulation~\eqref{eq:operator_L2}. This approach reduces the matrix dimension by a factor of two and is therefore more efficient, particularly in multiprecision arithmetic where both memory and runtime scale rapidly with precision. 
Eigenvalues are obtained using the \mcode{eigs} command with two selection modes depending on the problem type.
For the bright and dark solitons governed by the DNLS equation, the \mcode{'SM'} option (smallest magnitude) is used to extract eigenvalues closest to the origin, which correspond to internal or weakly unstable modes.
In contrast, for the parabolic reduction of the defocusing DNLS (Section~\ref{sec:parabolic}), the \mcode{'LA'} option (largest algebraic) is employed to compute the leading real eigenvalue associated with the translational mode.
In both cases, the solver tolerance is scaled to match the working precision, ensuring numerical stability and consistency across runs.

All numerical experiments are carried out on a workstation equipped with a 12th Gen Intel\textsuperscript{\textregistered} Core\texttrademark~i7--12700K CPU and 32~GB of RAM (31.7~GB usable). 
The number of grid points is varied from $N=500$ up to $N=65{,}250$, with the largest case corresponding to approximately 31.7~GB of memory. 
The multiprecision arithmetic is \textcolor{red}{typically set to 300~decimal digits (increased to 1000 for the most delicate runs)}, and the solver tolerances are chosen proportionally to the working precision. Our \textsc{Matlab} codes are provided in~\ref{app:code}.

\textcolor{red}{\subsection{Accuracy and Precision Requirements}\label{sec:accuracy}}

\textcolor{red}{Before presenting the results, we briefly analyze the error sources of the numerical implementation and discuss how they determine the precision and grid size needed in practice. Since the DNLS equation \eqref{eq:DNLS} is itself a lattice model, the spacing $h$ is a physical parameter rather than a discretization parameter; consequently, there is no spatial truncation error of the usual $\mathcal{O}(h^p)$ type. The error budget instead consists of three contributions.
(i)~\emph{Domain truncation.} The infinite lattice is truncated to $N$ sites with Neumann boundary conditions. Because the soliton tails approach their background values exponentially, $|\phi_j-\phi_\infty|\sim e^{-\kappa h|j|}$ with $\kappa=\mathcal{O}(\sqrt{|\omega|})$~\cite{Kevrekidis2009}, the boundary-induced perturbation of the eigenvalues is $\mathcal{O}(e^{-\kappa Nh/2})$. We therefore choose $N$ such that this quantity is negligible compared with the target eigenvalue scale $e^{-c/(2h)}$; this is the reason why $N$ must grow rapidly as $h$ decreases, cf.\ Table~\ref{tab:soliton-results}.
(ii)~\emph{Nonlinear-solver error.} The stationary states are computed by Newton iteration with the exact Jacobian \eqref{eq:Jdiscr} and tolerances set to the machine epsilon of the working precision, $10^{-\texttt{MP}}$; the quadratic convergence of Newton's method guarantees that the stationary profiles are accurate to essentially all \texttt{MP} digits.
(iii)~\emph{Roundoff in the eigenvalue computation.} This is the dominant and most restrictive error source. Standard backward-error results for eigenvalue problems~\cite{Higham2002,trefethen2020spectra} imply a perturbation of the squared operator \eqref{eq:operator_L2} of order $10^{-\texttt{MP}}\,\|L^{(h)}_-L^{(h)}_+\|$ in $\texttt{MP}$-digit arithmetic; the corresponding forward error in an eigenvalue also depends on its conditioning. Since the target eigenvalues are exponentially small, $\lambda^{2}\sim h^{-5}\mathrm{e}^{-c/h}$, with $c=\pi^{2}$ for bright and $c=\sqrt{2}\pi^{2}$ for dark solitons, the working precision must be chosen so that this backward-error level lies well below the target scale. This yields the practical scaling $\texttt{MP}\propto c/(h\ln 10)$ as $h$ decreases, with the proportionality margin determined empirically by convergence under increases of \texttt{MP}~\cite{BaileyBorwein2015}.}

\textcolor{red}{This requirement is borne out by our computations, as summarized in Table~\ref{tab:soliton-results}. For bright solitons the smallest resolved eigenvalue at $h=0.015$ is $|\lambda|\sim10^{-136}$, and $\texttt{MP}=300$ is found to be sufficient. For intersite dark solitons the exponential rate is larger by a factor $\sqrt2$, and at $h=0.0125$ the eigenvalue reaches $|\lambda|\sim10^{-236}$; there $\texttt{MP}=300$ no longer suffices and $\texttt{MP}=1000$ was required to obtain converged results. The onsite dark branch is limited by a different mechanism: for $h\lesssim0.9$ the relevant eigenvalues approach the continuous band, and increasing the working precision alone (we tested up to $\texttt{MP}=1000$) does not restore reliable results, since the difficulty lies in the spectral resolution of nearby modes rather than in roundoff. In practice we selected $\texttt{MP}$ for each run by increasing it until the computed eigenvalue no longer changed in the digits reported, and we verified in the same way that the results are insensitive to further increases of $N$. The mild deviations visible at the smallest $h$ in Figures~\ref{fig:spectrum_bright}(c,d) occur in precisely the regime where this precision requirement becomes most severe, indicating that they are of numerical origin rather than a genuine departure from the asymptotic prediction.}

\subsection{Bright Solitons}

\begin{figure}[htbp!]
	\centering
	\subfloat[Onsite bright soliton]{\includegraphics[scale=.5]{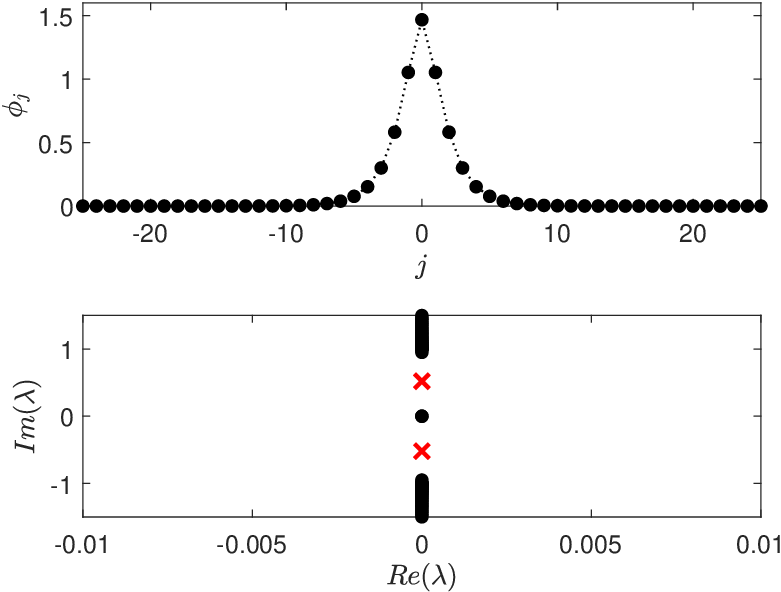}\label{subfig:prof_bright_onsite}}\quad
	\subfloat[Intersite bright soliton]{\includegraphics[scale=.5]{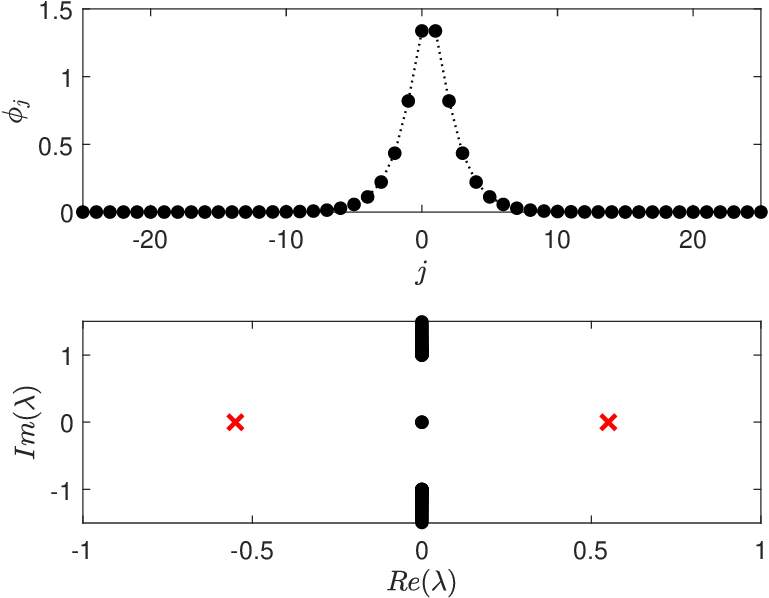}\label{subfig:prof_bright_intersite}}
	\caption{\textcolor{blue}{Profiles of discrete bright solitons obtained from the stationary DNLS equation~\eqref{eq:Fdiscr} with focusing nonlinearity $(\sigma=+1)$ and frequency $\omega=1$, together with the spectra of the corresponding linear operator \eqref{eq:Lpm_disc} in the complex plane; the symbols mark the eigenvalues of interest. Results are shown for $h=0.7$: (a)~onsite soliton; (b)~intersite soliton.}}
	\label{fig:prof_bright}
\end{figure}

\begin{figure}[thbp!]
	\centering
	\subfloat[Onsite bright soliton]{\includegraphics[scale=.5]{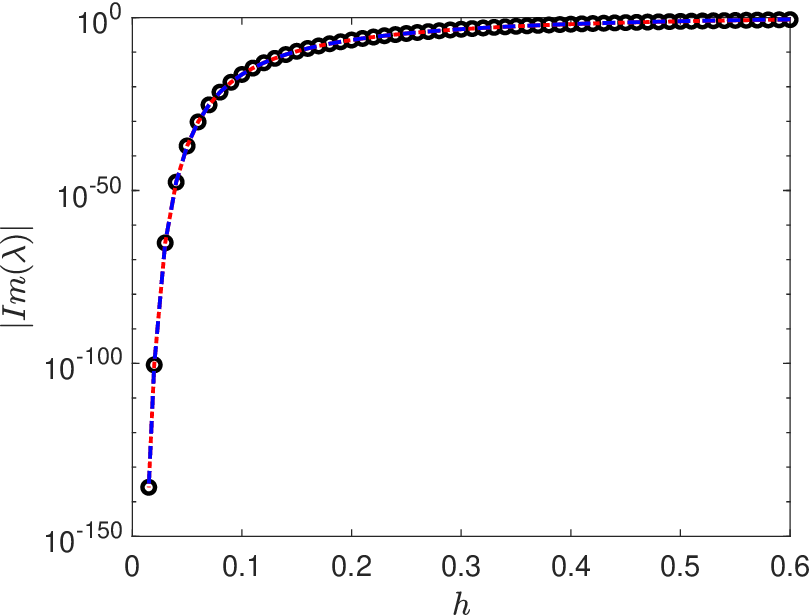}\label{subfig:spectrum_bright_onsite}}\quad
	\subfloat[Intersite bright soliton]{\includegraphics[scale=.5]{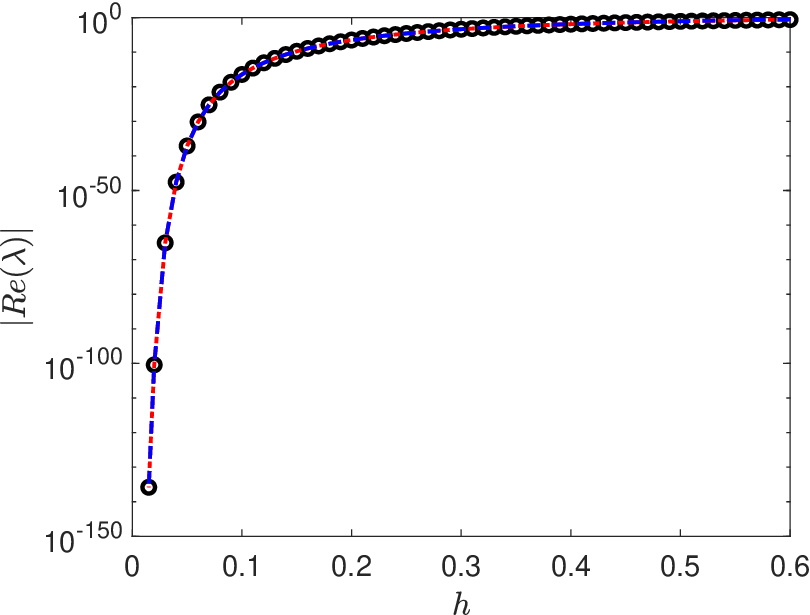}\label{subfig:spectrum_bright_intersite}}\\[1em]
	\subfloat[Onsite bright soliton (rescaled)]{\includegraphics[scale=.5]{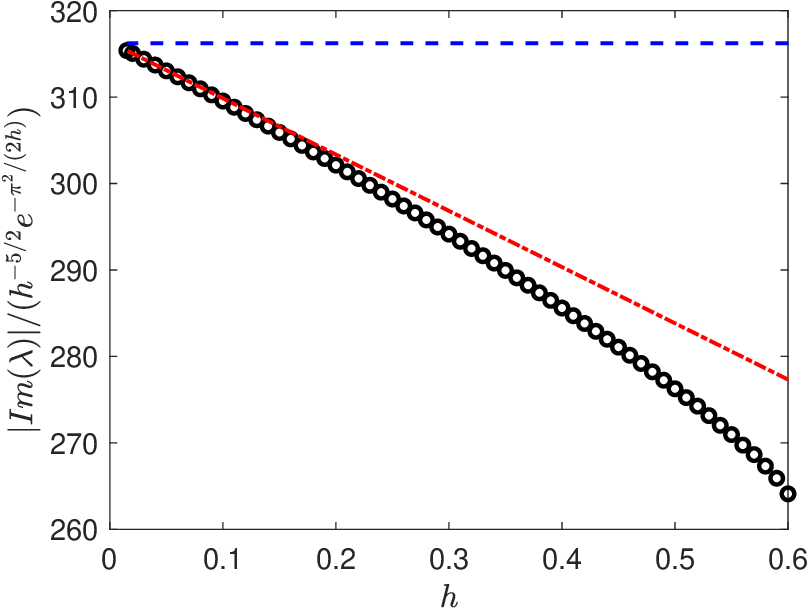}\label{subfig:Amp_spectrum_bright_onsite}}\quad
	\subfloat[Intersite bright soliton (rescaled)]{\includegraphics[scale=.5]{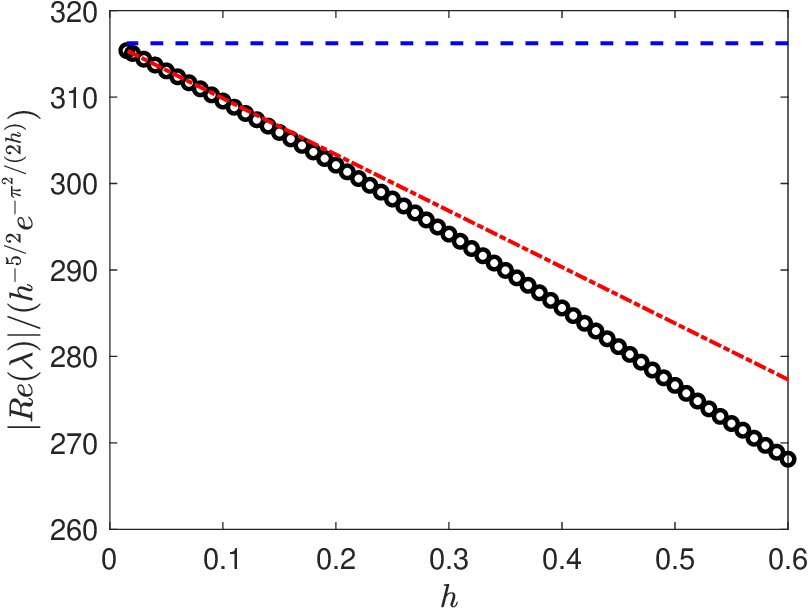}\label{subfig:Amp_spectrum_bright_intersite}}
	\caption{Spectra of discrete bright solitons. 
		Black open circles: numerical eigenvalues; 
		blue dashed lines: exponential asymptotics of Adriano \emph{et al.}~\cite{Adriano2025}; 
		red dash-dotted curves: higher-order corrections of Lustri \emph{et al.}~\cite{Lustri2025}. 
		(a) Onsite branch: $|\mathrm{Im}\,\lambda|$ versus $h$ on a logarithmic vertical scale. The internal mode remains purely imaginary and exponentially small as $h\to0$. 
		(b) Intersite branch: $|\mathrm{Re}(\lambda)|$ versus $h$ (log scale). A real pair $\pm\mathrm{Re}(\lambda)$ persists for all $h$, indicating a weak structural instability with exponentially small growth rate. 
		(c)-(d) Rescaled data: eigenvalues divided by $h^{-5/2}\,\mathrm{e}^{-\pi^{2}/(2h)}$, showing agreement with the predicted prefactors. 
	}
	\label{fig:spectrum_bright}
\end{figure}

For bright solitons, two asymptotic theories yield the same leading-order behavior as $h\to0$. 
The exponential asymptotics of Adriano \emph{et al.}~\cite{Adriano2025} predicts
\begin{equation}\label{eq:adriano_bright}
	\lambda^{2} \approx 4\pi\cdot2533\,h^{-5}\mathrm{e}^{-\pi^{2}/h}.
\end{equation}
This expression captures both the exponentially small dependence $\mathrm{e}^{-\pi^{2}/h}$ and the algebraic prefactor $h^{-5}$. 
Lustri \emph{et al.}~\cite{Lustri2025} further include the first higher-order correction for the eigenvalue itself,
\begin{equation}\label{eq:lustri_bright}
	|\lambda| \approx 316.355\left(1-\frac{\pi^{2}}{48}\,h\right)\,h^{-5/2}\mathrm{e}^{-\pi^{2}/(2h)}.
\end{equation}
Squaring~\eqref{eq:lustri_bright} reproduces~\eqref{eq:adriano_bright} up to $O(h)$ changes in the prefactor. 
Thus, both results agree on the exponentially small scale, with the distinction lying in the precision of the prefactor correction.

Figure~\ref{fig:spectrum_bright} confirms these predictions. 
For the onsite branch [panel~(a)], the internal mode lies on the imaginary axis. For the intersite branch [panel~(b)], the smallest eigenvalues form a real pair $\pm\mathrm{Re}(\lambda)$, persisting for all $h$. 
This indicates a structural instability with an exponentially small growth rate as $h\to0$. The numerical data for $\lambda$ closely follow the predicted exponential decay as a function of $h$, matching the asymptotic result of~\cite{Adriano2025}.

Panels~(c)-(d) show the same data rescaled by the leading exponentially small factor $h^{-5/2}\mathrm{e}^{-\pi^{2}/(2h)}$. 
In this representation, the asymptotics \eqref{eq:adriano_bright} will correspond to a horizontal line, while \eqref{eq:lustri_bright} corresponds to a straight line with a negative slope. The numerical onsite branch [panel~(c)] tends to the blue flat curve from \eqref{eq:adriano_bright}, and has the linear slope that is consistent with the $O(h)$ correction in~\eqref{eq:lustri_bright}. The intersite branch [panel~(d)] shows a similar trend, confirming the accuracy of the predicted prefactors. 
\textcolor{red}{Minor deviations at the smallest $h$ values are precision- and memory-limited (cf.\ Section~\ref{sec:accuracy}), the largest runs reaching $N=65{,}250$ and 31.7~GB of RAM.}

\begin{figure}[thbp!]
	\centering
	\subfloat[Onsite dark soliton, $\omega=-1$, $h=1.5$]{\includegraphics[scale=.5]{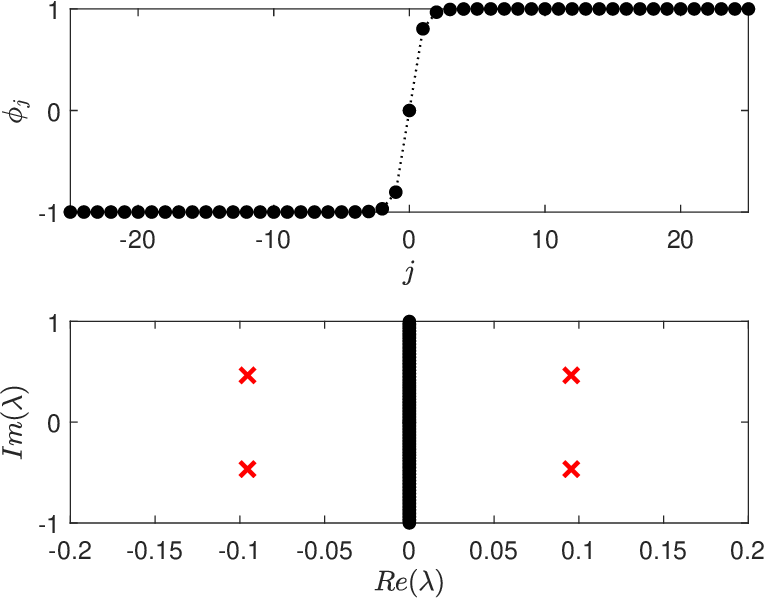}\label{subfig:prof_dark_onsite}}\quad
	\subfloat[Intersite dark soliton, $\omega=-1$, $h=0.7$]{\includegraphics[scale=.5]{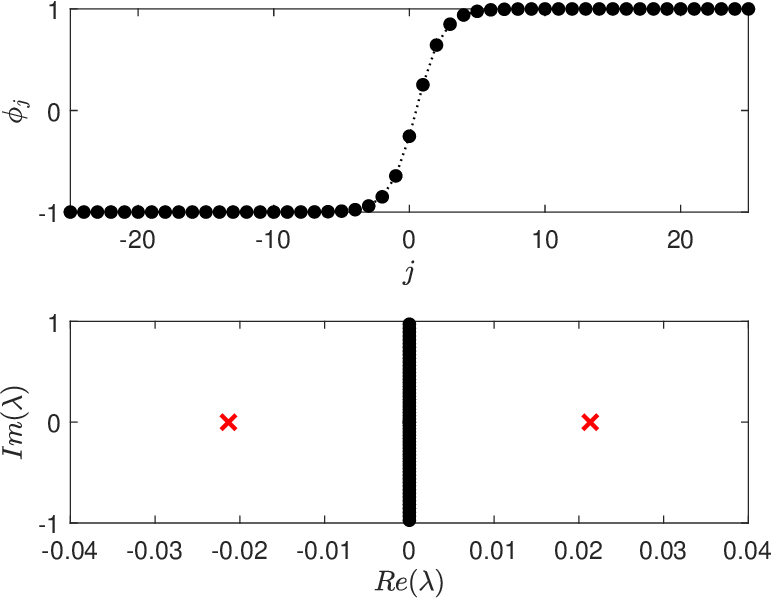}\label{subfig:prof_dark_intersite}}
	\caption{The same as Figure \ref{fig:prof_bright}, but for discrete dark solitons obtained from the stationary equation~\eqref{eq:Fdiscr} with defocusing nonlinearity $(\sigma=-1)$ and frequency $\omega=-1$. 
		(a) Onsite case ($h=1.5$); (b) intersite case ($h=0.7$).}
	\label{fig:prof_dark}
\end{figure}

\subsection{Dark Solitons}\label{sec:dark}

For the defocusing case $(\sigma=-1,\ \omega=-1)$, the stationary states shown in Figure~\ref{fig:prof_dark} correspond to kink-type (heteroclinic) dark solitons connecting the constant backgrounds $\phi \to \pm\sqrt{-\omega}$, in agreement with~\eqref{eq:dark}. 
In the onsite configuration [panel~(a)], the zero crossing of $\phi$ occurs on a lattice site, whereas in the intersite configuration [panel~(b)] it lies midway between two lattice points.

\begin{figure}[thbp!]
	\centering
	\subfloat[Onsite dark soliton]{\includegraphics[scale=.5]{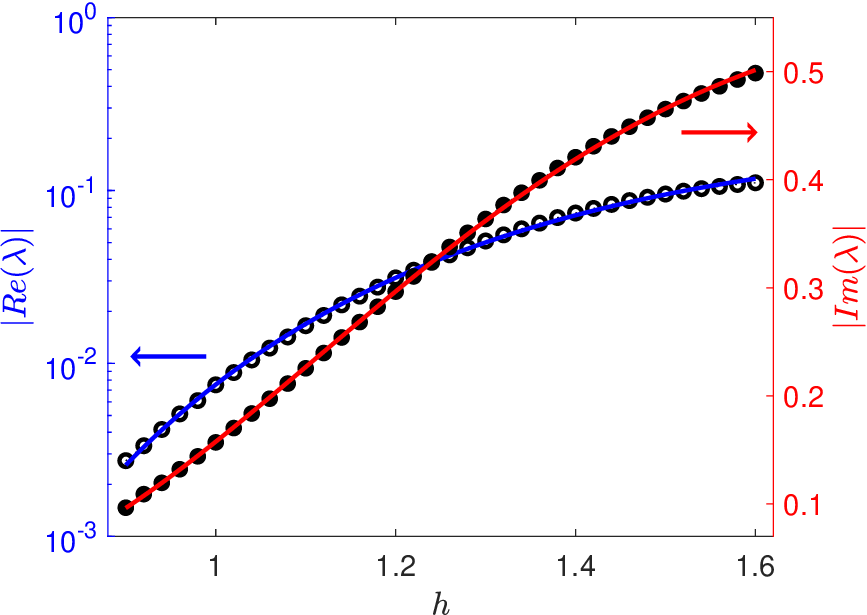}\label{subfig:spectrum_dark_onsite}}\quad
	\subfloat[Intersite dark soliton]{\includegraphics[scale=.5]{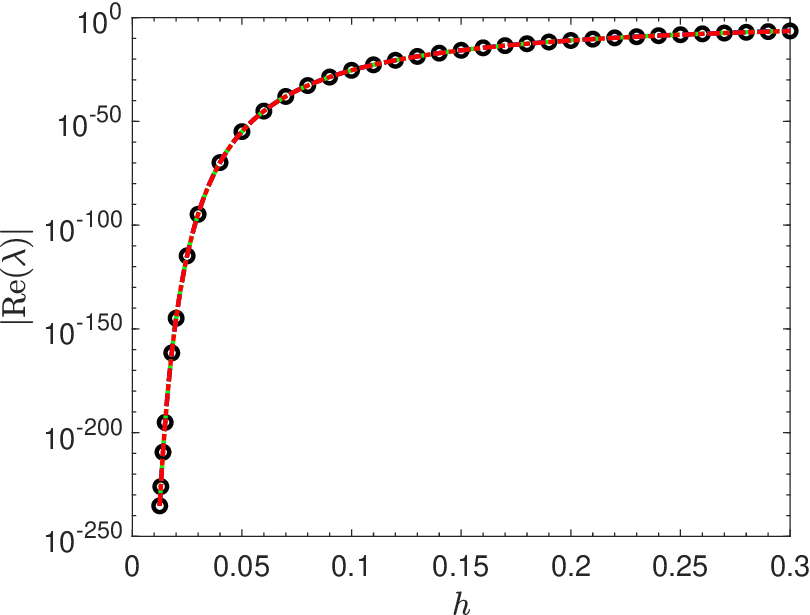}\label{subfig:spectrum_dark_intersite}}\\[1em]
	\subfloat[Intersite dark soliton (rescaled)]{\includegraphics[scale=.5]{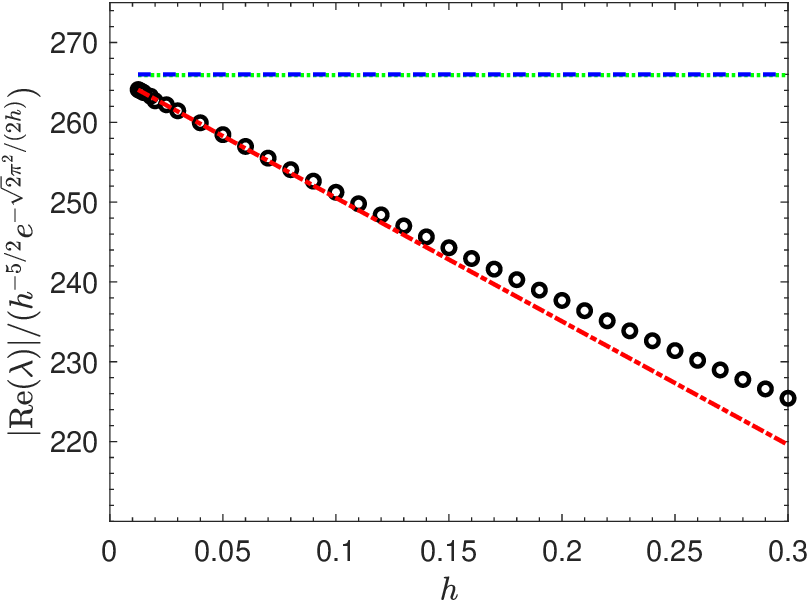}\label{subfig:Amp_spectrum_dark_intersite}}
	\caption{Spectra of discrete dark solitons \textcolor{red}{and comparison with asymptotic results.} 
		(a) Onsite branch: blue open circles show $\mathrm{Re}(\lambda)$ (left axis), and red dots show $\mathrm{Im}(\lambda)$ (right axis) versus $h$. 
		Green and cyan curves are exponential fits \textcolor{red}{of the form \eqref{eq:fit_onsite}} with parameters given in the text, confirming the exponentially small instability growth and corresponding prefactors. 
		\textcolor{red}{(b) Intersite branch: black open circles show $\mathrm{Re}(\lambda)$ versus $h$ on a logarithmic vertical scale. The green dotted curve is the factor-of-two-corrected prediction of Adriano \emph{et~al.}~\cite{Adriano2025}. {The blue dashed curve shows the leading-order term of the Lustri \emph{et~al.} formula, while the red dash-dotted curve shows the full next-order formula}~\cite{LustriDark2026}. {The two formulas are} given in \eqref{eq:adriano_dark} and \eqref{eq:lustri_dark}. 
		(c) Intersite branch (rescaled): the data and asymptotic curves from (b) are divided by $h^{-5/2}\exp[-\sqrt{2}\pi^{2}/(2h)]$ to isolate the prefactor. The green dotted line is the factor-of-two-corrected Adriano \emph{et~al.} value $265.913$, the blue dashed line is the {leading-order term $266.004$ of the} Lustri \emph{et~al.} formula, and the red dash-dotted line is the {full} next-order result $266.004-154.700h$.} Results extend down to the smallest $h$ accessible with available memory ($N=65{,}250$; 31.7~GB RAM).}
	\label{fig:spectrum_dark}
\end{figure}

The corresponding spectra in Figure~\ref{fig:spectrum_dark} indicate that both onsite and intersite dark solitons are spectrally unstable.  
Panel~(a) presents the onsite branch: blue open circles show $\mathrm{Re}(\lambda)$ (left axis), and red dots show $\mathrm{Im}(\lambda)$ (right axis) as functions of $h$.  The computation of the onsite branch becomes numerically delicate for $h\lesssim0.9$. \textcolor{red}{Even with $N=1000$ and working precisions raised as far as $1000$ digits}, the computed eigenvalues are sensitive to spectral collisions with the continuous band, which is expected in dark-soliton stability analysis \cite{johansson1999discreteness}.  

\textcolor{red}{Lustri \emph{et~al.}~\cite{LustriDark2026} derived onsite asymptotic formulas in the limit $h\to0$. However, our reliable onsite data are restricted to $h\geq0.9$, because for smaller $h$ the relevant eigenvalues undergo repeated interactions with the discretized continuous spectrum. The available data therefore do not reach a sufficiently small-$h$ regime for a robust quantitative comparison with those formulas. We consequently retain an empirical characterization of the onsite branch using the fit}
\begin{equation}\label{eq:fit_onsite}
	f(h)=c_{1}\,h^{-c_{2}}\,\mathrm{e}^{-c_{3}/h},
\end{equation}
using multiprecision data down to $h=0.9$.  
The best-fit parameters for the real part are $(c_{1},c_{2},c_{3})=(1.663\times10^{5},\,7.658,\,16.908)$, and for the imaginary part $(c_{1},c_{2},c_{3})=(8.986\times10^{2},\,4.438,\,8.648)$.  
The resulting green and cyan curves in Figure~\ref{fig:spectrum_dark}(a) reproduce both the exponential smallness and the algebraic prefactors over the full tested range of~$h$.

Panel~(b) reports the intersite branch.  All simulations extend to the smallest \textcolor{blue}{lattice spacing} permitted by available memory ($N=65{,}250$; 31.7~GB RAM). The black open circles represent the numerically computed values of $\mathrm{Re}(\lambda)$ on a logarithmic vertical scale.

{\color{red}
For the positive real eigenvalue of the intersite dark soliton, the exponential-asymptotic calculation of Adriano \emph{et~al.}~\cite{Adriano2025}, with the factor-of-two correction to the published expression (for which an erratum is in preparation), gives
\begin{equation}\label{eq:adriano_dark}
	|\lambda|_{\mathrm{A}} \approx \sqrt{2\sqrt2\pi^{2}\!\cdot2533}\,
	h^{-5/2}\mathrm{e}^{-\sqrt2\pi^{2}/(2h)}
	\approx 265.913\,h^{-5/2}\mathrm{e}^{-\sqrt2\pi^{2}/(2h)}.
\end{equation}
Lustri \emph{et~al.}~\cite{LustriDark2026} independently obtained the {approximation}
\begin{equation}\label{eq:lustri_dark}
	|\lambda| \approx (266.004-154.700h)\,h^{-5/2}\mathrm{e}^{-\sqrt2\pi^{2}/(2h)},
\end{equation}
as $h\to0$. Figure~\ref{fig:spectrum_dark}(b) compares {the two} formulas directly with the multiprecision eigenvalues. After the factor-of-two correction, the leading prefactor $265.913$ {in \eqref{eq:adriano_dark}} is within $0.04\%$ of the leading-order term $266.004$ {in the expression \eqref{eq:lustri_dark}; the two predictions} approach the numerical results as $h$ decreases. {These} analytical expressions reproduce the variation over more than two hundred orders of magnitude, while the {next-order formula by Lustri \emph{et~al.}~\eqref{eq:lustri_dark}} captures the finite-$h$ trend most accurately.

Panel~(c) shows the same comparison after division by $h^{-5/2}\exp[-\sqrt2\pi^{2}/(2h)]$. The corrected prediction of Adriano \emph{et~al.} becomes the horizontal line $265.913$, the {leading-order term of the Lustri \emph{et~al.} formula} becomes the nearly coincident horizontal line $266.004$, and the {full next-order formula} becomes the straight line $266.004-154.700h$. The numerical prefactor approaches $\sim263$ at the smallest accessible spacings, showing good quantitative agreement with {the leading-order terms of the two formulas}. Because the {full next-order formula} already describes the first finite-$h$ trend, an additional empirical fit is neither needed nor used for the intersite branch.
}

\textcolor{red}{We note that the eigenvalue problem~\eqref{eq:bdg_disc} underlying this comparison is gapless, a structure that in general complicates the standard exponential-matching arguments (see, e.g., \cite{pelinovsky2008stability,Adriano2025Maxwell}). It is therefore worth testing the same asymptotic method on a related problem in which a spectral gap is present, which is the subject of the next subsection.}

\begin{figure}[thbp!]
	\centering
	\subfloat[Intersite dark soliton]{%
		\includegraphics[scale=.5]{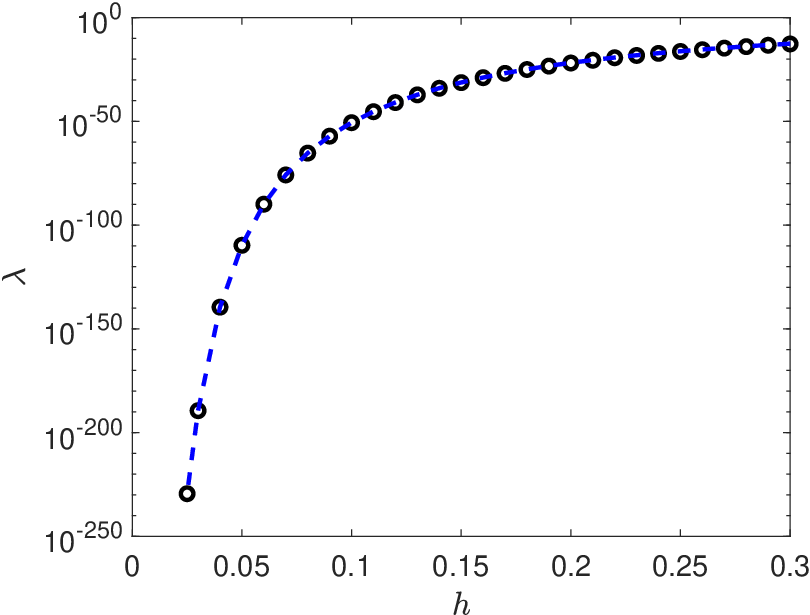}%
		\label{subfig:spectrum_dark_parabolic}}%
	\quad
	\subfloat[Intersite dark soliton (rescaled)]{%
		\includegraphics[scale=.5]{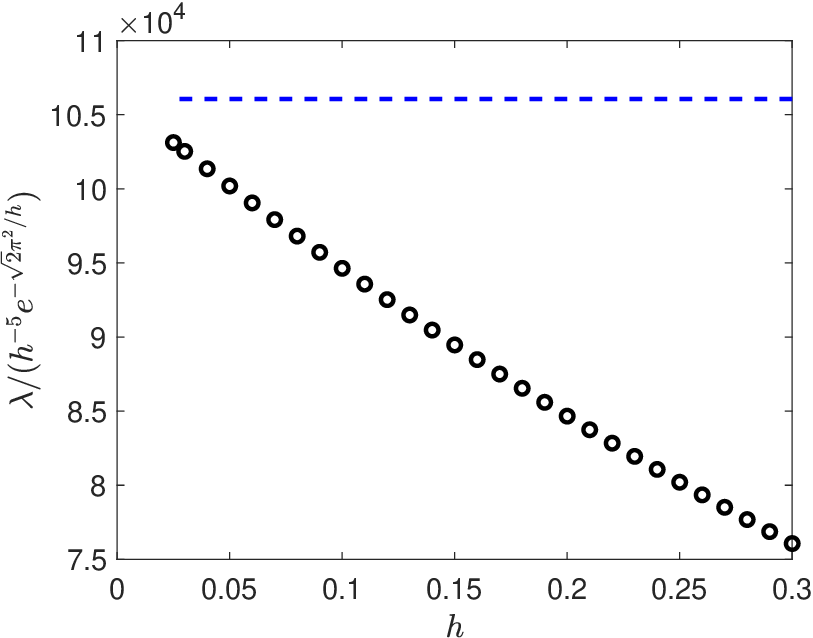}%
		\label{subfig:Amp_spectrum_dark_parabolic}}%
	\caption{Spectra from the corresponding parabolic \textcolor{red}{(gradient-flow)} problem for the intersite dark soliton.
		\textcolor{red}{(a) Magnitude $|\lambda|$ of the negative intersite eigenvalue of $-L_+^{(h)}$ versus lattice spacing $h$ on a logarithmic vertical scale.} 
		\textcolor{red}{(b) Same magnitude rescaled by $h^{-5}\exp(-\sqrt{2}\pi^{2}/h)$ to isolate the prefactor. 
		Black open circles: multiprecision numerical data for $|\lambda|$; 
		dashed blue line: magnitude of the asymptotic prediction, $3\sqrt{2}\pi^{2}\!\cdot\!2533 \approx 1.056\times10^{5}$, from Eq.~\eqref{eqn:eigval approx phi4} in~\ref{app:parabolic}.} 
		The limiting value of the numerical results confirms the predicted exponential scaling and algebraic dependence. 
		Simulations extend to the smallest $h$ feasible within available resources ($N=20{,}000$, \texttt{MP}~$=300$). 
		The finest run at $h=0.025$ required approximately $1.03\times10^{4}$\,s ($\approx171.6$\,min) of wall-clock time.}
	\label{fig:spectrum_parabolic}
\end{figure}

\subsection{Parabolic Problem}\label{sec:parabolic}

\textcolor{red}{Having compared the factor-of-two-corrected Adriano and the Lustri exponential-asymptotic prefactors for intersite dark solitons in Section~\ref{sec:dark}, we now perform an independent cross-check using a different, related problem. We consider an auxiliary problem that involves the \emph{same} stationary dark-soliton profiles, but whose linearization operator possesses a spectral gap---unlike the gapless operator~\eqref{eq:bdg_disc} of the full DNLS eigenvalue problem: the parabolic (gradient-flow) counterpart of the defocusing DNLS equation, described in \ref{app:parabolic}. If the same exponential-asymptotic method used in~\cite{Adriano2025} also agrees with the multiprecision numerics for this second, gapped problem, this further reinforces confidence in the combined multiprecision-plus-exponential-asymptotics approach across different classes of eigenvalue problems. This subsection is therefore an integral part of our validation of the exponential asymptotics, not a digression. The parabolic problem considered here is a discrete Allen--Cahn (real Ginzburg--Landau) equation. Although the cited works of Styles and co-workers address different phase-field equations rather than the present eigenvalue problem, they provide representative theorem-based numerical analyses: Barrett, N\"urnberg, and Styles~\cite{BarrettNurnbergStyles2004} prove stability and convergence for a finite-element approximation of a degenerate Cahn--Hilliard system, while Deckelnick, Elliott, and Styles~\cite{DeckelnickElliottStyles2016} establish $\Gamma$-convergence and finite-element convergence for a double-obstacle phase-field formulation of an inverse problem. These works therefore illustrate the rigorous validation paradigm suggested by the reviewer, rather than serving as direct validations of our DNLS computations.}

We now consider the discrete eigenvalue problem
\begin{equation}\label{eq:phi4eig}
	\textcolor{red}{-L_+^{(h)}u = \lambda u,}
\end{equation}
evaluated about the previously computed intersite dark soliton.
\textcolor{red}{Its negative eigenvalue closest to the origin corresponds to the weak translational mode associated with the dark-soliton kink. The numerical routine computes the corresponding positive largest-algebraic eigenvalue $\mu$ of $L_+^{(h)}$; hence $\lambda=-\mu<0$, and Figure~\ref{fig:spectrum_parabolic} displays the magnitude $|\lambda|=\mu$.}
Unlike the DNLS case, this auxiliary problem focuses directly on the leading real eigenvalue, which captures the exponentially small breaking of translational symmetry.
An asymptotic approximation of this eigenvalue is derived in~\ref{app:parabolic}.

Figure~\ref{fig:spectrum_parabolic} shows the multiprecision results for the intersite dark soliton obtained from~\eqref{eq:phi4eig}.
\textcolor{red}{Panel~(a) presents $|\lambda|$, the magnitude of the negative intersite eigenvalue of $-L_+^{(h)}$, as a function of the lattice spacing~$h$ on a logarithmic scale.
Black open circles indicate the numerical magnitudes, while the blue dashed curve represents the magnitude of the asymptotic prediction \eqref{eqn:eigval approx phi4}.
The observed exponential decay of~$|\lambda|$ as $h \to 0$ confirms the presence of an exponentially weak translational mode caused by discrete symmetry breaking.}

\textcolor{red}{Panel~(b) displays the same magnitude data rescaled by the leading asymptotic factor $h^{-5}\exp(-\sqrt{2}\pi^{2}/h)$ to isolate the prefactor.}
The numerical results agree quantitatively with the asymptotic expression~\eqref{eqn:eigval approx phi4}, confirming the expected exponential scaling.
\textcolor{red}{This agreement, obtained independently for a gapped operator, is consistent with the good quantitative agreement already found for the gapless DNLS operator~\eqref{eq:bdg_disc} in Section~\ref{sec:dark}, and together the two results reinforce confidence in the exponential-asymptotic method of~\cite{Adriano2025} across both classes of eigenvalue problems.}

\subsection{\textcolor{red}{Computational Cost and Performance}}
\begin{table}[ht]
	\centering
	\caption{Summary of eigenvalue computations for bright and dark solitons. 
		The table reports the parameter $h$, number of grid points $N$, number of eigenvalues $k$ computed with \texttt{eigs}, 
		\textcolor{red}{the relevant nonzero spectral value, with the parabolic entries reported as the magnitude $|\lambda|$ of the negative intersite eigenvalue of $-L_+^{(h)}$}, runtime in seconds, and the multiprecision setting (\texttt{MP}). 
	}
	\label{tab:soliton-results}
	
	\centering
	\begin{tabular}{lllrrlrr}
		\toprule
		Soliton & Type      &\multicolumn{1}{c}{$h$} & \multicolumn{1}{c}{$N$} & $k$ & \multicolumn{1}{c}{$\lambda$} & Runtime (s) & \texttt{MP} \\
		\midrule
		\multirow{8}{*}{Bright}
		& \multirow{4}{*}{Onsite}
		& 0.6   &   100  &   2     & $0.25378\,i$          &   0.20262   & 300 \\
		&       & 0.05   & 10000  &   2     & $7.67378\times10^{-38}\,i$         &   6.25735   & 300 \\
		&       & 0.02   & 40000  &   2     & $3.87195\times10^{-101}\,i$       &  33.13658   & 300 \\
		&       & 0.015  & 65250  &   2     & $1.51860\times10^{-136}\,i$       &  65.36363   & 300 \\
		\cmidrule(lr){2-8}
		& \multirow{4}{*}{Intersite}
		& 0.6   &   100  &   2     & $0.25764$          &  0.17789  & 300 \\
		&       & 0.05   & 10000  &   2     & $7.67378\times10^{-38}$         &   5.46921   & 300 \\
		&       & 0.02   & 40000  &   2     & $3.87195\times10^{-101}$       &  30.24384   & 300 \\
		&       & 0.015  & 65250  &   2     & $1.51860\times10^{-136}$       &  45.23069   & 300 \\
		\midrule
		\multirow{12}{*}{Dark}
		& \multirow{4}{*}{Onsite}
		& 1.6   &   200  &  32     & $0.11086 + 0.49863\,i$  &  15.68691   & 300 \\
		&       & 1.5   &  1000  & 160     & $0.09542 + 0.46534\,i$ &  97.63927   & 300 \\
		&       & 1.0   &  1000  & 160     & $0.00752 + 0.15696\,i$ &  87.95271   & 300 \\
		&       & 0.9   &  1000  & 160     & $0.00274 + 0.09650\,i$ & 104.06226   & 300 \\
		\cmidrule(lr){2-8}
		& \multirow{4}{*}{Intersite}
		& 0.3   &   200  &   2     & \textcolor{blue}{$3.60789\times10^{-7}$}           &   0.17821   & 300 \\
		&       & 0.05  & 10000  &   2     & \textcolor{blue}{$1.11509\times10^{-55}$}        &   7.92786   & 300 \\
		&       & 0.02  & 40000  &   2     & \textcolor{blue}{$1.32668\times10^{-145}$}       &  33.60314   & 300 \\
		&       & 0.0125& 65250  &   2     & \textcolor{blue}{$5.11587\times10^{-236}$}       & 123.15882   & 1000 \\
		\cmidrule(lr){2-8}
		& \multirow{4}{*}{Parabolic}
		&  0.3     &    200    &    1     &     $1.94852\times10^{-13} $  &      11.90993      &  300   \\
		&       &    0.15   &    1000     &   1    &      $4.56497\times10^{-32} $      &  94.29642 &300   \\
		&       &    0.05   &    10000     &   1    &      $1.86505\times10^{-110} $      &  $1.55\times10^3$ &300   \\
		&       &    0.025   &    20000     &   1    &      $3.57328\times10^{-230} $      &  $1.03\times10^4$ &300   \\
		\bottomrule
	\end{tabular}
\end{table}

Table~\ref{tab:soliton-results} illustrates our eigenvalue computations. We report representative values of the lattice spacing~$h$, the number of grid points~$N$, the number of eigenvalues~$k$ computed using the squared operator formulation, \textcolor{red}{the relevant nonzero spectral value (reported as $|\lambda|$ for the parabolic problem)}, the wall-clock runtime, and the multiprecision setting (\texttt{MP}).

\textcolor{red}{In the $\lambda$ column}, the listed values correspond to different spectral characteristics: for bright-onsite solitons, $\lambda$ denotes 
the purely imaginary internal mode (excluding the neutral phase mode at $\lambda=0$); for bright-intersite and dark-intersite solitons, it represents the positive real growth rate of the unstable pair; for dark-onsite solitons, it shows the complex pair $\mathrm{Re}(\lambda) + i\mathrm{Im}(\lambda)$; \textcolor{red}{and for the parabolic problem, the positive tabulated value is $|\lambda|$, the magnitude of the negative intersite eigenvalue of $-L_+^{(h)}$.}

As the lattice spacing~$h$ decreases, the bright-onsite eigenvalue rapidly approaches zero, confirming spectral stability with purely imaginary modes.
By contrast, the bright-intersite branch retains a real pair for all~$h$, indicating a weak but persistent structural instability with an exponentially small growth rate.
For both bright branches, computing only two eigenvalues ($k=2$) is sufficient, since the relevant modes are well isolated near the origin.
The runtime scales moderately with~$N$ and remains manageable even for the largest grid ($N=65{,}250$ at \texttt{MP}$=300$), which required about one minute and 31.7~GB of memory.

For dark solitons, both onsite and intersite configurations exhibit nonzero real parts of~$\lambda$, confirming spectral instability.
The onsite branch is numerically delicate because the unstable mode lies close to the continuous spectrum.
To capture this mode accurately, we computed several eigenvalues around the origin (up to $k=160$ for $N=1000$) and employed higher precision for smaller~$h$.
\textcolor{red}{Even when the working precision is raised to \texttt{MP}$=1000$, results for $h\lesssim0.9$ remain unreliable, owing to interactions between discrete and continuous spectral components; the entries listed in the table therefore stop at $h=0.9$.}
In contrast, the intersite dark branch is easier to resolve spectrally, requiring only $k=2$, although its smallest eigenvalues are extremely small.
At the largest~$N$, \texttt{MP}$=1000$ was necessary, which increased the runtime to approximately two minutes.

\textcolor{red}{For the parabolic problem, the positive largest-algebraic eigenvalue $\mu$ of~$L_+^{(h)}$ (see Eq.~\eqref{eq:Lpm_disc}) was computed to capture the translational mode of the discrete front; equivalently, the corresponding eigenvalue of $-L_+^{(h)}$ is $\lambda=-\mu<0$.}
\textcolor{red}{The numerical magnitudes agree quantitatively with the magnitude of the asymptotic prediction in~\ref{app:parabolic} (Eq.~\eqref{eqn:eigval approx phi4}), showing the same exponential scaling $|\lambda| \sim h^{-5}\exp(-\sqrt{2}\pi^{2}/h)$.}
At the smallest spacing ($h=0.025$), the computation required $N=20{,}000$ grid points, \texttt{MP}$=300$, and a runtime of approximately $1.03\times10^{4}$~s.
For small~$h$, the main challenge arises from the rapid growth of~$N$ needed to resolve the front and from the exponentially small magnitude of~$\lambda$, both of which demand high precision and substantially increase computational cost.
\textcolor{red}{These results confirm that the parabolic reduction successfully reproduces the exponentially weak translational spectral splitting associated with the same stationary dark-soliton profiles.}

\section{Conclusions}\label{sec:conclusion}
{
	We have carried out a systematic multiprecision study of stationary states and spectral stability in the discrete nonlinear Schr\"odinger equation, and compared the numerical results directly with exponential asymptotics. The main point of this work is that exponentially small eigenvalues, which determine soliton stability, cannot be resolved in standard double precision. To compute them reliably, the numerical precision must match their asymptotic scale.}

{
	For bright solitons, the numerical results agree very well with the analytical predictions. The onsite branch remains spectrally stable, while the intersite branch has an exponentially small real instability. The computed eigenvalues follow the predicted scaling laws, including the refined prefactor corrections. In this case, the relevant eigenvalues are well separated from the rest of the spectrum, so only a small number of eigenvalues is needed. The comparison between analysis and computation is therefore clear and quantitative.}

{
	Dark solitons show a more delicate behavior. Their unstable eigenvalues lie close to the continuous spectrum, which makes the problem more sensitive to discretization and numerical precision. To obtain convergence, both higher precision and a larger subset of eigenvalues are required. \textcolor{red}{For the intersite branch, the factor-of-two-corrected formula of Adriano \emph{et~al.}~\cite{Adriano2025} has prefactor $265.913$, essentially coincident with the leading prefactor $266.004$ of Lustri \emph{et~al.}~\cite{LustriDark2026}. The multiprecision computations approach $\sim263$ and agree quantitatively with {the leading-order terms of both formulas}, while the {full Lustri formula} captures the finite-$h$ trend. For the onsite branch, by contrast, spectral interactions limit the reliable computations to $h\geq0.9$, outside the small-$h$ regime needed for a quantitative test of the corresponding asymptotics, so that we retain only an empirical fit over the accessible range. The intersite agreement is further corroborated by an independent cross-check using a related parabolic reduction, where a spectral gap is present and the numerical results likewise match the asymptotic prediction.}}

{
	The main limitation of multiprecision computation is computational cost. Exponentially small spectral effects require fine grids and high working precision, which significantly increase memory usage and runtime. In our largest DNLS simulations ($N=65{,}250$), memory usage reached approximately 31.7~GB. \textcolor{red}{For the parabolic problem, computations at the finest lattice spacing required runtimes of several hours even at \texttt{MP}$=300$.} These limitations are mainly due to hardware constraints, not to weaknesses of the numerical formulation.}

{
	The computational framework presented in~\ref{app:code}, which combines Newton-type solvers with exact Jacobians and a squared-operator formulation for spectral analysis, provides a reproducible and flexible approach. Although developed here for the DNLS equation, the same methodology can be applied to other lattice dynamical systems and nonlinear wave problems where stability is controlled by exponentially small spectral effects.}

{\textcolor{red}{Our results also point to several directions for future work.
First, a rigorous numerical analysis of multiprecision eigenvalue computations, with proven error bounds in the spirit of theorem-based analyses for phase-field discretizations~\cite{BarrettNurnbergStyles2004,DeckelnickElliottStyles2016}, would put the error estimates of Section~\ref{sec:accuracy} on a firm theoretical footing. Second, it would be interesting to extend the present framework to other classes of solutions, such as discrete breathers and solutions on a finite background~\cite{AkhmedievAnkiewicz1997}, whose stability spectra are expected to pose similar precision challenges.}}

{
	In conclusion, multiprecision arithmetic is not simply a numerical improvement, but a necessary tool for validating beyond-all-orders asymptotic results in discrete Hamiltonian systems. By combining high-precision computation with exponential asymptotics, we build a clear and consistent link between analysis and numerical experiments in regimes where standard precision is insufficient.}

\section*{CRediT authorship contribution statement}
{\textbf{Rudy Kusdiantara}: Writing -- review \& editing, Writing -- original draft, Validation, Software, Formal analysis, Conceptualization; \textbf{Farrell T. Adriano}: Validation, Methodology, Formal analysis; \textbf{Hadi Susanto}: Writing -- review \& editing, Writing -- original draft, Resources, Methodology, Formal analysis, Conceptualization.}

\section*{Declaration of generative AI and AI-assisted technologies in the writing process}

During the preparation of this work, the authors utilized Grammarly and ChatGPT to enhance language and readability. After using these tools/services, the authors reviewed and edited the content as needed and take full responsibility for the content of the publication.

\section*{Declaration of competing interest}
{The authors declare that they have no known competing financial interests or personal relationships that could have appeared to influence the work reported in this paper.}

\section*{Acknowledgment}
{R.K. acknowledges financial support from the Ministry of Higher Education, Science, and Technology of the Republic of Indonesia and the Indonesia Endowment Fund for Education (LPDP) through the Strategic Research Downstreaming Funding Program under the Industry Invitation and Synergy Scheme (Pendanaan Program Hilirisasi Riset Strategis Skema Ajakan Industri dan Sinergi).} HS acknowledges support by Khalifa University through a Research \& Innovation Grant under project ID KU-INT-RIG-2024-8474000789.

\section*{Data availability}
No data was used for the research described in the article.

\bibliographystyle{elsarticle-num}
\bibliography{references}



	
	
	

\newpage
\appendix
\section{Code and Reproducibility Notes}\label{app:code}

This appendix documents the \textsc{Matlab} programs used in our computations.  
We tested with \textsc{Matlab} R2024b and the Advanpix Multiprecision Computing Toolbox v5.4.0.16035 (see Sec.~\ref{sec:numerics}).  
Each block below shows (i) the task, (ii) parameters to modify, and (iii) the relevant code.

\subsection{Environment and precision}\label{app:env}
\begin{tabular}{lcl}
	\textbf{Task}&: & Load Advanpix and set the working precision. \\[4pt]
	\textbf{User edits}&: & Adjust the \texttt{addpath} to match your installation and choose the number of digits \texttt{MP}. \\
\end{tabular}
\begin{lstlisting}
	%% DNLS stationary state (multiprecision fsolve)
	% Compute a stationary profile u and its linearized spectrum.
	
	clc; clearvars; close all; tic
	
	% -- Advanpix path (edit to local installation) ----------------------
	addpath('...\Multiprecision Computing Toolbox\');
	
	% -- Precision -------------------------------------------------------
	MP = 300;                   % choose 50--300 depending on difficulty
	mp.Digits(MP);
	
	% ---------------- Parameters ----------------------------------------
	n0    = mp(0);              % center: n0=0 (onsite), n0=1/2 (intersite)
	omega = mp(-1);             % frequency: omega>0 bright; omega<0 dark
	sigma = mp(-1);             % nonlinearity: +1 focusing, -1 defocusing
	N     = 2000;               % grid points (large N = finer grid)
	h     = mp('0.15');         % lattice spacing
	
	% -- Discrete Laplacian with Neumann BC ------------------------------
	D = sparse(diag(-2*ones(1,N)) + diag(ones(1,N-1),1) + diag(ones(1,N-1),-1));
	D(1,1) = -1; D(N,N) = -1;   % Neumann correction
	Dh2 = (mp(1)/h^2) * mp(D);  % scaled Laplacian
	
	% Grid index (for initial guess and plotting)
	n = mp((-N/2+1:N/2)).';
	
	% ---------------- Initial guess -------------------------------------
	% Bright soliton (focusing): uncomment if sigma=+1, omega>0
	% u0 = sqrt(2*omega) .* sech(h*(n - n0).*sqrt(omega));
	
	% Dark soliton (defocusing): default here (sigma=-1, omega<0)
	u0 = sqrt(-omega) .* tanh(h*(n - n0).*sqrt(-omega/2));
	
	% ---------------- Nonlinear solver (trust-region) -------------------
	options = optimset('Display','iter', ...
	'TolFun', mp('eps'), ...
	'TolX', mp('eps'), ...
	'Algorithm','trust-region-dogleg', ...
	'Jacobian','on', ...
	'MaxIter', 100);
	
	% Residual and Jacobian handle
	FJ = @(u) dnls_resjac_mp(u, Dh2, omega, sigma);
	
	% Solve nonlinear system
	[u, ~, exitflag, output] = fsolve(@(x) FJ(x), u0, options);
	
	% Rescue if convergence fails
	if exitflag <= 0
	kick = mp(10)^(-MP) * (2*rand(mp(size(u))) - 1);
	[u, ~, exitflag, output] = fsolve(@(x) FJ(x), u + kick, options); %#ok<ASGLU>
	end
	
	% ---------------- Linear operators ---------------------------------
	Lminus = Dh2 + spdiags(-omega + sigma*(u.^2), 0, N, N);
	Lplus  = Dh2 + spdiags(-omega + 3*sigma*(u.^2), 0, N, N);
	
	% Squared operator for spectrum
	J = Lminus * Lplus;
	
	% Eigenpairs of J near zero or largest algebraic (lambda^2 ~ -eig(J))
	k_small = 2; 
	opts.tol = mp('eps');
	
	% --- Choose eigs option manually ---
	% Use 'SM' for bright/dark solitons (DNLS)
	% Use 'LA' for parabolic reduction
	eigs_option = 'SM';    % change to 'LA' when solving the parabolic problem
	
	% Compute eigenvalues
	temp_eig = eigs(J, k_small, eigs_option, opts);
	lambda   = sqrt(temp_eig);
	
	
	fprintf('h = %.6g,  |lambda(2)| = %.12g\n', double(h), double(lambda(2)));
	duration = toc;
\end{lstlisting}

Residual and Jacobian function:
\begin{lstlisting}
	function [F,J] = dnls_resjac_mp(u, Dh2, omega, sigma)
	% Residual and Jacobian for stationary DNLS
	F = -omega*u + Dh2*u + sigma*(u.^3);
	if nargout > 1
	N = length(u);
	J = Dh2 + spdiags(-omega + 3*sigma*(u.^2),0,N,N);
	end
	end
\end{lstlisting}

\subsection{Quick tuning guide}\label{app:tune}
\begin{itemize}
	\item \textbf{Precision:} set \texttt{MP} (50--150 typical; 300 for hardest cases).  
	\item \textbf{Grid:} choose \texttt{N} (500--65{,}250) and spacing \texttt{h} (\(h=L/N\)); make \(L\) large enough so soliton tails decay.  
	\item \textbf{Model:} set \(\sigma=\pm1\) and the sign of \(\omega\) (\(\omega>0\) for bright, \(\omega<0\) for dark).  
	\item \textbf{Solver:} tighten \texttt{TolFun}/\texttt{TolX} with higher \texttt{MP}; increase \texttt{MaxIter} if necessary.  
	\item \textbf{Spectrum:} use the squared operator for robustness; raise \texttt{k\_small} to capture more internal modes.  
	\item \textbf{Eigenvalue option:} set the \texttt{eigs} selection mode according to the problem type:  
	\begin{itemize}
		\item \texttt{'SM'} --- for bright and dark solitons in the DNLS equation (computes eigenvalues nearest to zero).  
		\item \texttt{'LA'} --- for the parabolic reduction (computes the largest algebraic eigenvalue corresponding to the translational mode).  
	\end{itemize}
\end{itemize}

\subsection{Troubleshooting}\label{app:trouble}
\begin{itemize}
	\item \textbf{No convergence:} improve the initial guess, increase \texttt{MaxIter}, or add a small random perturbation (\texttt{kick}).  
	\item \textbf{Complex eigenvalues near zero:} increase \texttt{MP}, reduce \texttt{opts.tol}, or apply shift-invert.  
	\item \textbf{Memory exhausted:} reduce \texttt{N} or \texttt{MP}; the largest case (\(N=65{,}250\)) requires about 31.7~GB of RAM.  
	\item \textbf{Missing or inaccurate eigenvalues:} if expected eigenvalues are not captured, adjust the \texttt{eigs} selection mode.  
	Use \texttt{'SM'} for bright and dark solitons (to target eigenvalues near zero), or switch to \texttt{'LA'} for the parabolic problem (to extract the largest algebraic eigenvalue).  
\end{itemize}

\section{Eigenvalues of the Parabolic Problem}\label{app:parabolic}

This appendix presents the linear stability analysis of the parabolic analogue of the defocusing DNLS equation,
\begin{equation} \label{eqn:phi4}
	\phi_{t} = \frac{\phi_{j+1}-2\phi_{j}+\phi_{j-1}}{h^{2}} + \phi_{j} - \phi_{j}^{3}.
\end{equation}
The stationary solutions of~\eqref{eqn:phi4} coincide with the dark-soliton (kink) profiles of the defocusing DNLS.  
As $h \to 0$, the stationary front solution of~\eqref{eqn:phi4} admits the asymptotic expansion
\begin{equation} \label{eqn:phi asymp}
	\phi(x) \sim \tilde{\phi}(x) + R_{N}(x),
\end{equation}
where $\tilde{\phi}(x) = \tanh\!\big((x - x_{0})/\sqrt{2}\big)$ is the leading-order (continuum) kink,  
$x = h n$ is the slow spatial variable, $x_{0} = h n_{0}$ denotes the kink center with $n_{0} \in \{0,\,1/2\}$, and $R_{N}(x)$ is an exponentially small remainder satisfying $R_{N}(x) \to 0$ as $x \to -\infty$.  
As $x \to \infty$,
\begin{equation}
	R_{N}(x) \sim -\frac{\pi \cdot 2533}{4}\,h^{-4}\,e^{-\sqrt{2}\pi^{2}/h}
	\sin\!\left(\frac{2\pi(x - x_{0})}{h}\right)e^{\sqrt{2}x}.
\end{equation}
Eliminating this exponentially growing tail restricts the admissible locations of the kink center to  
$n_{0}=0$ (onsite solution) or $n_{0}=1/2$ (intersite solution), modulo integer shifts.

Linearization around the stationary solution~$\phi$ gives the discrete eigenvalue problem~\eqref{eq:phi4eig},  
where $L_{+}^{(h)}$ is defined as in~\eqref{eq:Lpm_disc} with $\omega=-1$ and $\sigma=-1$.  
To obtain an asymptotic approximation of the bifurcating zero eigenvalue of $-L_{+}^{(h)}$ from the continuum limit, we expand the corresponding eigenfunction in powers of~$\lambda$:
\begin{equation} \label{eqn:eigenfunction expansion}
	u = u_{0} + \lambda u_{1} + \cdots,
\end{equation}
where the leading-order term $u_{0}$ is taken as the translational eigenmode in the continuum limit,
\[
u_{0} = \phi'(x).
\]
From~\eqref{eqn:phi asymp}, $u_{0}$ possesses an exponentially growing tail as $x \to \infty$:
\begin{equation}
	u_{0} \sim -\frac{\pi^{2} \cdot 2533}{2}\,h^{-5}\,e^{-\sqrt{2}\pi^{2}/h}
	\cos\!\left(\frac{2\pi(x - x_{0})}{h}\right)e^{\sqrt{2}x}.
\end{equation}
This unbounded term must be cancelled by the next-order correction.  
At $\mathcal{O}(\lambda)$, the governing equation is
\begin{equation}
	-L_{+}^{(h)}u_{1} = u_{0}.
\end{equation}
In the limit $h \to 0$, $L_{+}^{(h)}$ reduces to its continuum counterpart $L_{+}$, and asymptotically
\begin{equation}
	u_{1} \sim \nu(x)\,\tilde{\phi}'(x),
\end{equation}
where
\begin{equation}
	\nu(x) \sim \frac{1}{16\sqrt{2}}\!\int_{-\infty}^{\infty}\!(\tilde{\phi}'(s))^{2}\,\mathrm{d}s\; e^{2\sqrt{2}x}
	= \frac{1}{6\sqrt{2}}\,e^{2\sqrt{2}x}, \qquad \text{as } x \to \infty.
\end{equation}
Hence,
\begin{equation}
	u_{1} \sim \frac{1}{6\sqrt{2}}\,e^{\sqrt{2}x}, \qquad x \to \infty.
\end{equation}
Including both terms in~\eqref{eqn:eigenfunction expansion}, we obtain the large-$x$ behaviour
\begin{equation}
	u \sim u_{0} + \lambda u_{1}
	\sim \left[-\frac{\pi^{2} \cdot 2533}{2}\,h^{-5}\,e^{-\sqrt{2}\pi^{2}/h}
	\cos\!\left(\frac{2\pi(x - x_{0})}{h}\right)
	+ \frac{\lambda}{6\sqrt{2}}\right] e^{\sqrt{2}x}.
\end{equation}
Eliminating this exponentially growing tail yields the asymptotic expression for the eigenvalue:
\begin{equation} \label{eqn:eigval approx phi4}
	\lambda \sim 3\sqrt{2}\,\pi^{2}\!\cdot\!2533\,h^{-5}\,e^{-\sqrt{2}\pi^{2}/h}\cos(2\pi n_{0}),
	\qquad h \to 0.
\end{equation}
Equation~\eqref{eqn:eigval approx phi4} implies that the zero eigenvalue in the continuum limit bifurcates into a positive eigenvalue for the onsite solution and a negative one for the intersite solution.  
Both bifurcating eigenvalues are exponentially small in~$h$.  
{Our result is identical to that} derived using the modulation approach of King and Chapman~\cite{KING_CHAPMAN_2001}, who obtained {it through} the slow dynamics of the kink center $x_{0}(t)$.

\end{document}